\documentclass[11pt]{article}
\usepackage[english]{babel}
\usepackage{fullpage}
\usepackage{graphics,graphicx}
\usepackage{float}
\usepackage{wrapfig}
\usepackage{amssymb}
\usepackage{amsmath}
\usepackage{amsthm}

\theoremstyle{plain}
\newtheorem{theorem}{Theorem}
\newtheorem{definition}[theorem]{Definition}
\newtheorem{proposition}[theorem]{Proposition}

\title{On Determining if Tree-based Networks Contain Fixed Trees}

\renewcommand\footnotemark{}   

\author{Maria Anaya$^{1}$,
Olga Anipchenko-Ulaj$^{2}$,
Aisha Ashfaq$^{1}$,
Joyce Chiu$^{3}$,\\
Mahedi Kaiser$^{4}$,
Max Shoji Ohsawa$^{3}$,
Megan Owen$^{4}$ $^*$,
Ella Pavlechko$^{5}$, \\
Katherine St.~John$^{4,7}$,
Shivam Suleria$^{3}$,
Keith Thompson$^{6}$, and
Corrine Yap$^{5}$
\thanks{$^{1}$Queensborough Community College, City University of New York (CUNY);
$^{2}$City College of New York, CUNY;
$^{3}$Brooklyn College, CUNY;
$^{4}$Department of Mathematics \& Computer Science, Lehman College, CUNY, Bronx, NY 10468;
$^{5}$Sarah Lawrence College;
$^{6}$College of Staten Island, CUNY;
$^{7}$Division of Invertebrate Zoology, American Museum of Natural History, New York, NY 10024;
$^*$Corresponding author: Megan Owen ({\tt\small megan.owen@lehman.cuny.edu})}
}

\begin{document}
\maketitle

\begin{abstract}
    We address an open question of Francis and Steel about phylogenetic networks and trees.  They give a polynomial time algorithm to decide if a phylogenetic network, $N$, is tree-based and pose the problem:  given a fixed tree $T$ and network $N$, is $N$ based on $T$? We show that it is $NP$-hard to decide, by reduction from $3$-Dimensional Matching (3DM), and further, that the problem is fixed parameter tractable.
\end{abstract}


\section{Introduction}

\begin{figure}[t]
    \centering
    \begin{tabular}{cccc}    
 \includegraphics[height = 1in]{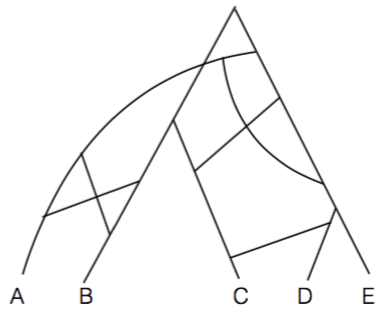} &
    \includegraphics[height = 1in]{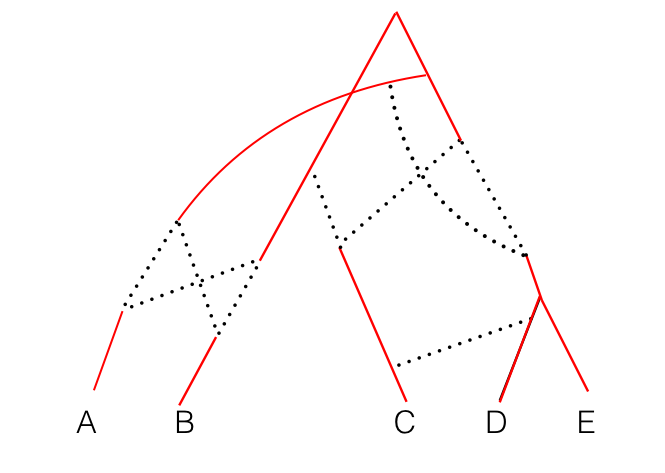} &
    \includegraphics[height = 1in]{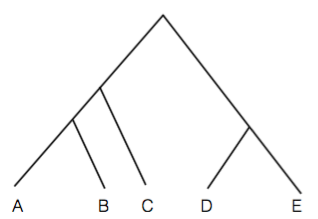} &
    \includegraphics[height=1in]{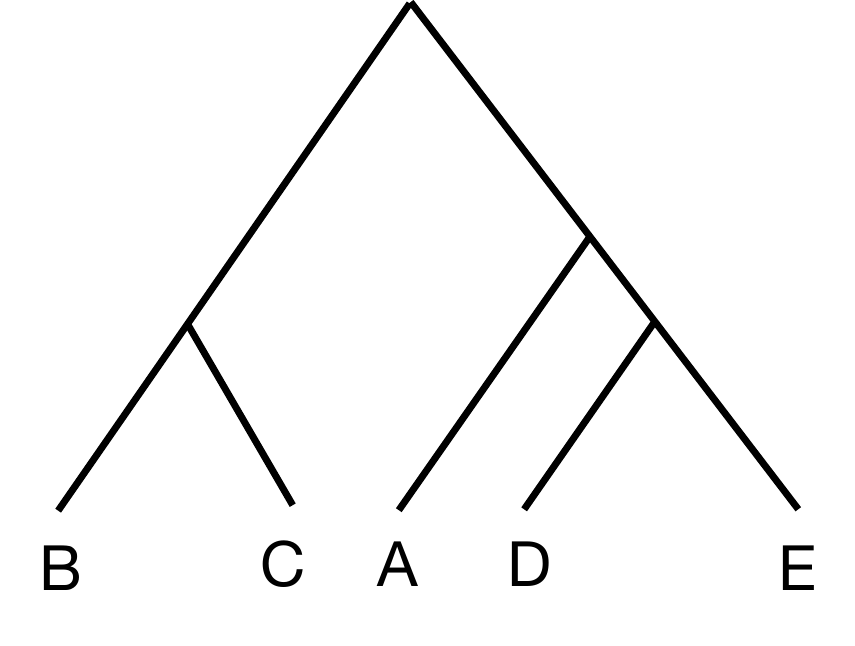} 
        \\ a) & b) & c) & d)
    \end{tabular}
    \caption{a)  A phylogenetic network on leaf set
    ${\cal L} = \{A,B,C,D,E\}$.
    b)  The same network after applying the $\lambda$ and $Y$ reduction rules.  After this process, the dotted edges are still unresolved.  Their resolution yields different possible trees on which the network could be based.
    c)  A possible tree, but one on which $N$ is not based.
    d)  A tree on which $N$ is based.
    }    
    \label{fig:treeNetwork}
\end{figure}

A canonical question in biology is to determine the evolutionary history of a set of species.  These histories are often represented by trees or directed acyclic graphs, the latter of which are called phylogenetic networks.  Networks add both extra flexibility and extra complexity for modeling evolution.  It is computationally hard to decide if a general phylogenetic network displays a tree \cite{kanj2008} but computationally tractable for some special classes of networks \cite{vanIersel2010}. 
Francis and Steel \cite{francis2015} introduced a new class of networks that captures much of the flexibility of general networks but has nice properties that make this class easier to use. Roughly, a phylogenetic network is {\em tree-based} if there exists an induced subtree on the same leaf set for which all non-tree edges are adjacent to tree edges (see Figure~\ref{fig:treeNetwork}).  Francis and Steel show that determining if a phylogenetic network is tree-based can be done in polynomial time via a reduction to $2$-SAT.  In their paper, they pose three open problems about tree-based phylogenetic networks: (i) how to characterize the set of rooted binary phylogenetic trees on which a tree-based network could be based; (ii) if it is possible to determine if a network is based on a given tree; and (iii) if there is a network on leaf set ${\cal L}$ which is tree-based for all trees on ${\cal L}$.  The first question was addressed by Semple \cite{semple2015}, Zhang \cite{zhang2015}, and Jetten and van Iersel \cite{jetten2016}. The third was answered affirmatively by Hayamizu and Fukumizu \cite{hayamizu2015}.
In this paper, we address the second problem of the difficulty in deciding if a given network $N$ is based on a specific tree $T$.  Surprisingly, we show that this is NP-hard via a reduction from $3$-Dimensional Matching (3DM).  Furthermore, we show that the problem is fixed parameter tractable.

\section{Definitions and Background}

We first provide the underlying definitions and then restate useful reduction rules from Francis and Steel \cite{francis2015} that we use to show the fixed parameter tractability of the problem.
We follow the definitions of \cite{francis2015} unless otherwise noted.

\begin{definition}
A {\em binary phylogenetic network} over leaf-set ${\cal L}$ is any directed acyclic graph $N = (V, A)$ for which:
\begin{itemize}
    \item ${\cal L} \subset V$ and each $l \in {\cal L}$, called a \emph{leaf}, has out-degree $0$ and in-degree $1$;
    \item There is a unique vertex of in-degree $0$, called the {\em root} (denoted $\rho$), which has out-degree $1$ or $2$; and 
    \item Every vertex other than $\rho$ or a leaf has either in-degree $2$ and out-degree $1$ or in-degree $1$ and out-degree $2$.
\end{itemize}
\end{definition}

\begin{definition}
$N$ is a {\em tree-based network} with base tree $T$ if $N$ can be described as follows:  
\begin{itemize}
    \item Subdivide each arc of $T$ as many times as required, calling the resulting degree-2 vertices
    {\em attachment points} 
    and the resulting tree $T'$ a {\em support tree for $N$ derived from $T$}.
    \item Next, sequentially place additional {\em linking arcs} between pairs of attachment points, such that the network remains binary.
\end{itemize}
See Figure~\ref{fig:treeNetwork} for an example of a tree-based network with base tree. 
\end{definition}

Francis and Steel \cite{francis2015} show that it takes polynomial time to decide if a network is tree-based by 
applying several useful reductions to the initial network.  While we show the question of determining if, given a tree $T$ and network $N$, $N$ is based on $T$ is NP-hard, we use these helpful reductions in showing the fixed parameter tractability of our problem.

\begin{proposition} (Proposition 3 of \cite{francis2015}):  Consider a binary phylogenetic network $N$ over leaf set ${\cal L}$.
\begin{enumerate}
    \item If each vertex of $N$ of in-degree 2 has parents of out-degree 2, then $N$ is tree-based.
    \item If $N$ has a vertex of in-degree 2 whose parents both have out-degree 1, then $N$ is not tree-based.
\end{enumerate}
\end{proposition}
Due to their shape, we refer to the vertices of in-degree 2 as {\em $Y$-vertices} and those with out-degree 2 as {\em $\lambda$-vertices}.  Note that a rooted, binary network with $m$ leaves must have at least $m-2$ $\lambda$-vertices (not including the root vertex, which we assume here to have out-degree 2), corresponding to the branching of the underlying tree, and the number of $Y$-vertices is exactly the difference between $m-2$ and the number of $\lambda$-vertices.  The latter quantity forms the parameter for the fixed parameter tractability result in Section~\ref{sec:results}.

\begin{figure}[t]
    \centering
    \begin{tabular}{clclclc}
        \includegraphics[width=.75in]{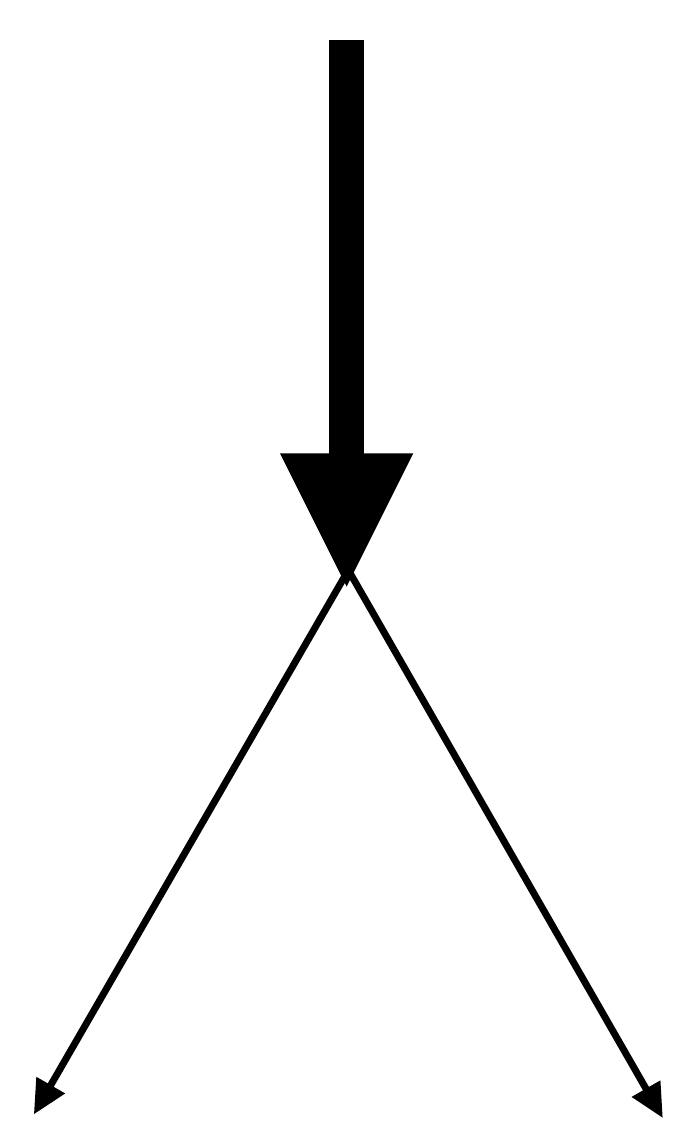} && 
        \includegraphics[width=.75in]{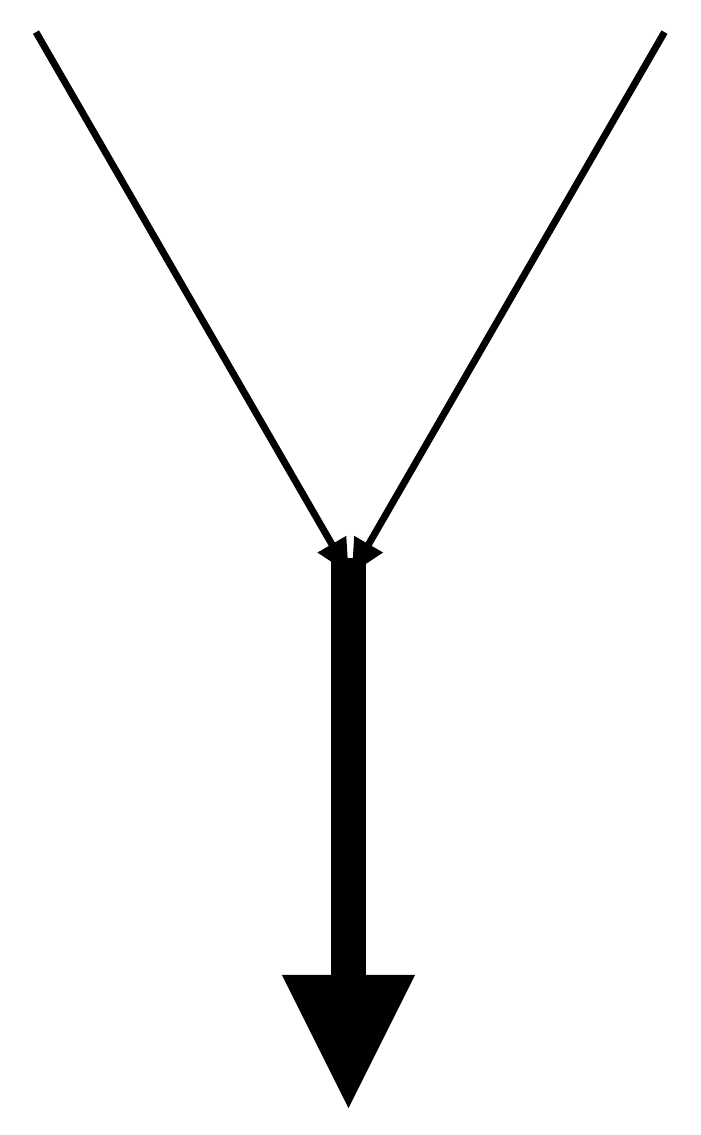} &&
        \includegraphics[width=.75in]{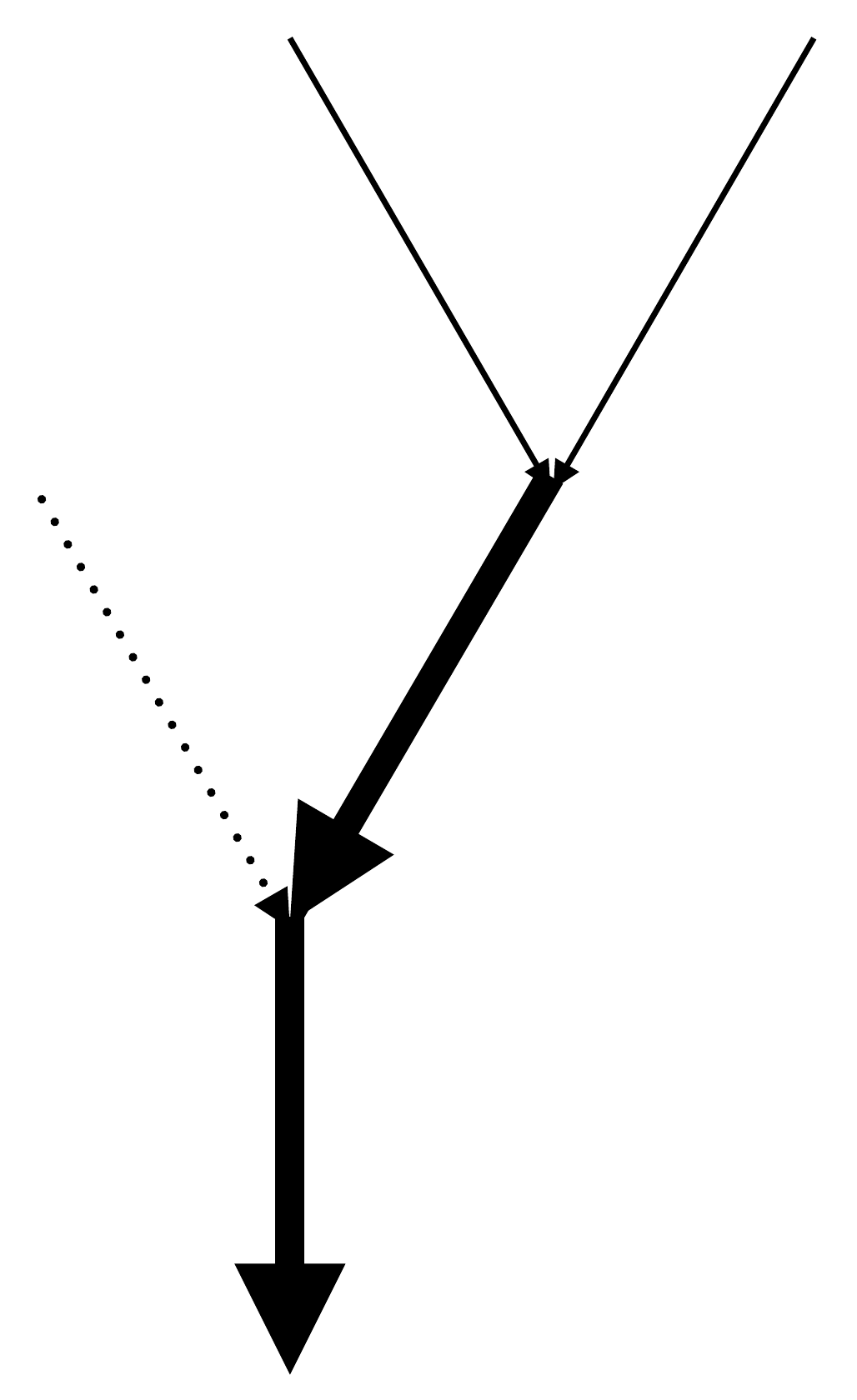} && 
        \includegraphics[width=.75in]{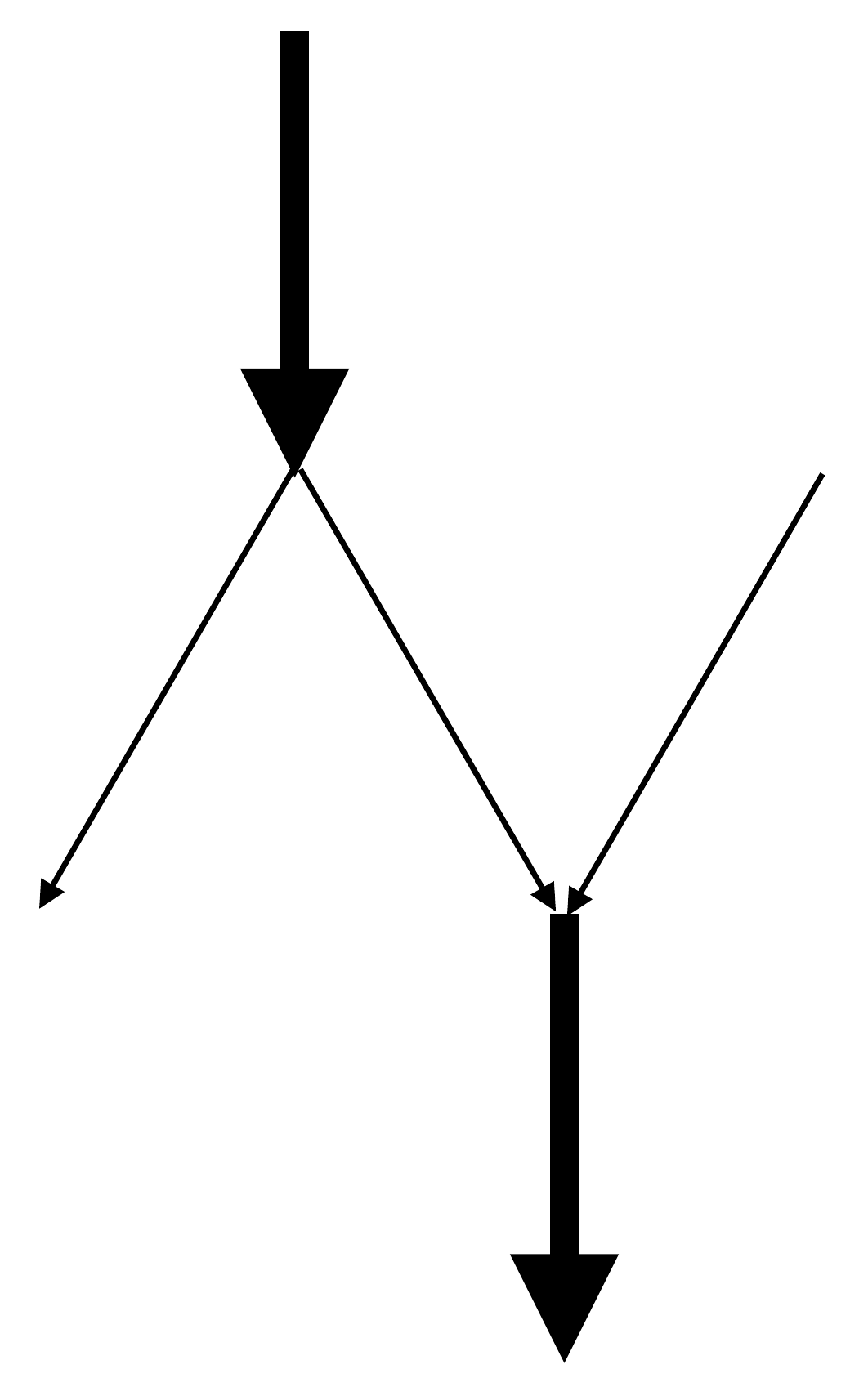}\\
    a) &\mbox{\hspace{.275in}} & 
    b) &\mbox{\hspace{.275in}} & 
    c) &\mbox{\hspace{.275in}} & 
    d) 
    \end{tabular}
    \caption{a) An example of a $\lambda$-vertex:  The bold incoming edge must be part of the underlying tree.
    b) An example of a $Y$-vertex:  The bold outgoing edge must be part of the underlying tree.
    c) 
    Similarly when two $Y$-vertices are in series, then the lower one is resolved (the non-tree edge is dotted), and
    d) $\lambda$- and $Y$-vertices that cannot be resolved without more information.
    }
    
    \label{fig:lambdaY}
\end{figure}

The algorithm from \cite{francis2015} to determine if a network as tree-based has two reductions that can be applied repeatedly to simplify the problem (see Figure~\ref{fig:lambdaY}).  Using the terminology above, they are:
\begin{itemize}
    \item {\em $Y$-Reduction:}  The outgoing edge of a $Y$-vertex must be in the tree.  This follows since one incoming edge must be a non-tree edge, and the remaining two edges must be in the tree.
    \item {\em $\lambda$-Reduction:}  The incoming edge of a $\lambda$-vertex must be in the tree.  If not, then there would be no (directed) path from the root to the vertex.  This cannot happen, since all vertices in the network occur in the tree (only  edges are removed to yield the tree).
\end{itemize}

\section{Results}
\label{sec:results}

We address the second question of Francis and Steel \cite{francis2015}:
\begin{quote}
{\em
Given a tree-based network $N$ and an arbitrary rooted binary phylogenetic tree $T$, can it be decided in polynomial time whether or not $N$ is based on $T$?
}
\end{quote}
Our first result is that it is NP-hard to decide this question.
Our second result is to provide a fixed parameter tractable algorithm, parametrized by the number of $Y$-vertices in the network.

\subsection{NP-Hardness}

We can reduce our problem from 3-dimensional matching (3DM), one of Karp's 21 NP-complete problems \cite{karp1972} ({\tt SP1} in \cite{garey1979}):  

\begin{samepage}
\begin{quote}
{\sc 
3 Dimensional Matching (3DM)}\\
{\sc Input:}  The sets $X$, $Y$, $Z$ with $|X|=|Y|=|Z| = n$
and $k$ subsets $S=\{S_1,\ldots,S_k\}$ where $S_i = (x_i, y_i, z_i)$ for some $x_i \in X$, $y_i \in Y$ and $z_i \in Z$.\\
{\sc Question:}  Is there a matching $M \subseteq S$ such that $|M| = n$ and $\bigcup M = X \times Y \times Z$?
\end{quote}
\end{samepage}

\begin{figure}[t]
    \centering
    \includegraphics[width=2.75in]{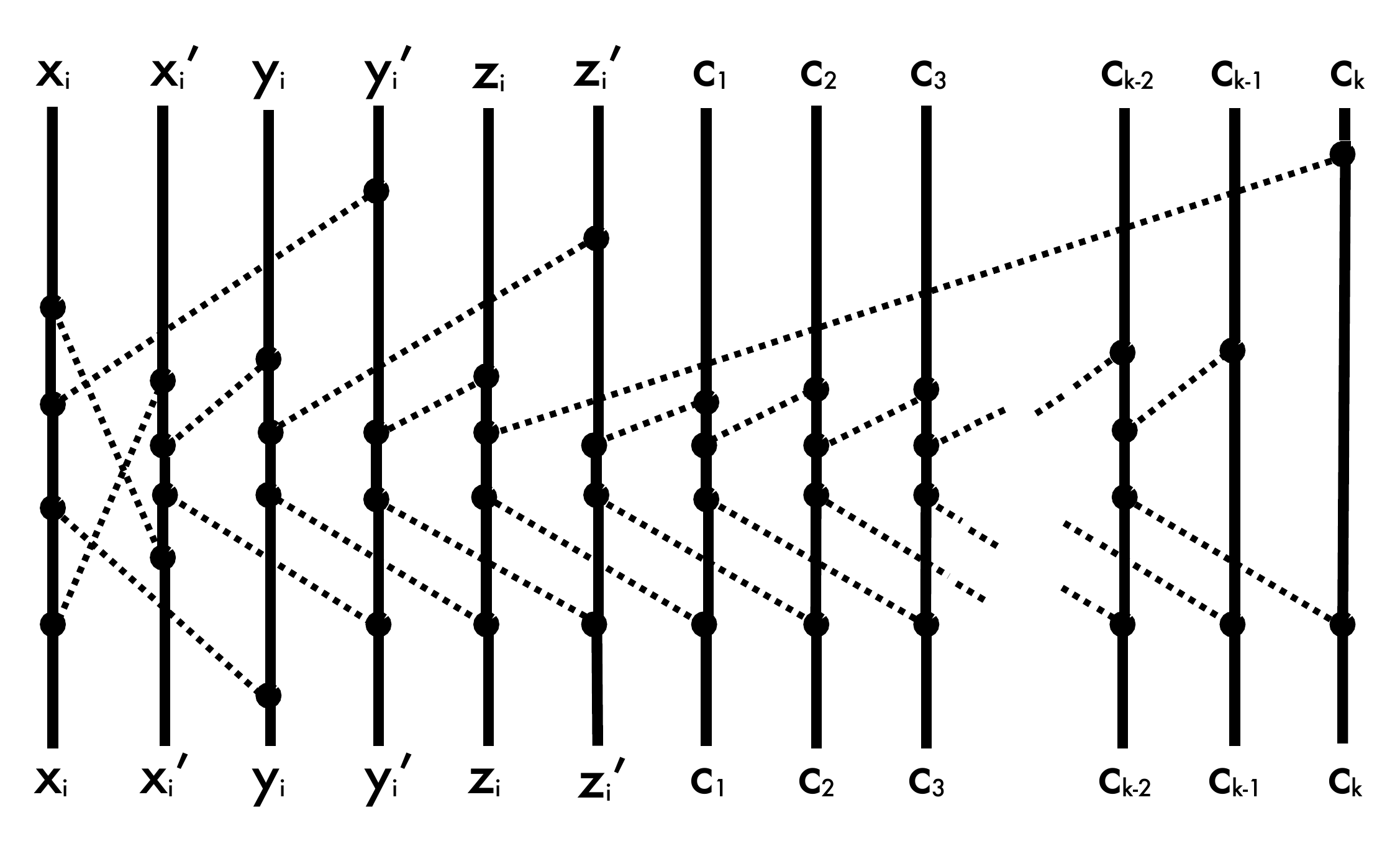}
    \hspace{.5in}
    \includegraphics[width=2.75in]{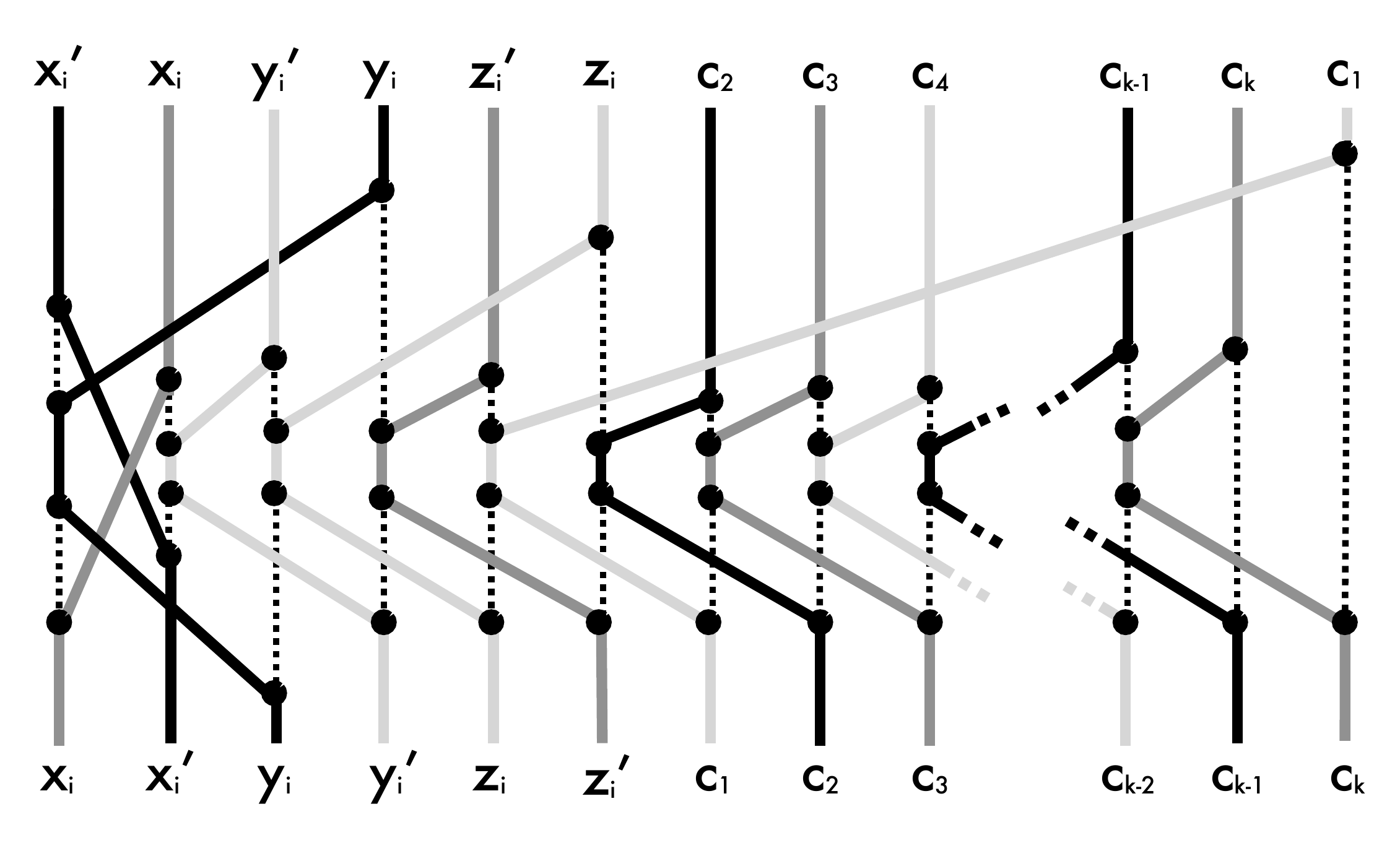}
    \caption{The gadget encoding the triple $S_i = (x_i,y_i,z_i)$.  This gadget uses six of the element ``threads'' ($x_i,x_i',y_i,y_i',z_i,z_i'$) as well as all $k$ counter threads (represented here as $c_1, \ldots, c_k$) which keep track of how many triples are used in the matching. By construction, there are exactly two ways to traverse this gadget such that the network stays tree-based.  In the left picture, no element nor counter thread changes position, and this corresponds to $S_i$ not being part of the matching.  The right picture corresponds to $S_i$ being part of the matching.  Note that the $k$ counter threads ``shift'' only when the $S_i$ is part of the matching.}.  
    \label{fig:gadget}
\end{figure}

\begin{figure}[t]
    \centering
    \includegraphics[height=2.5in]{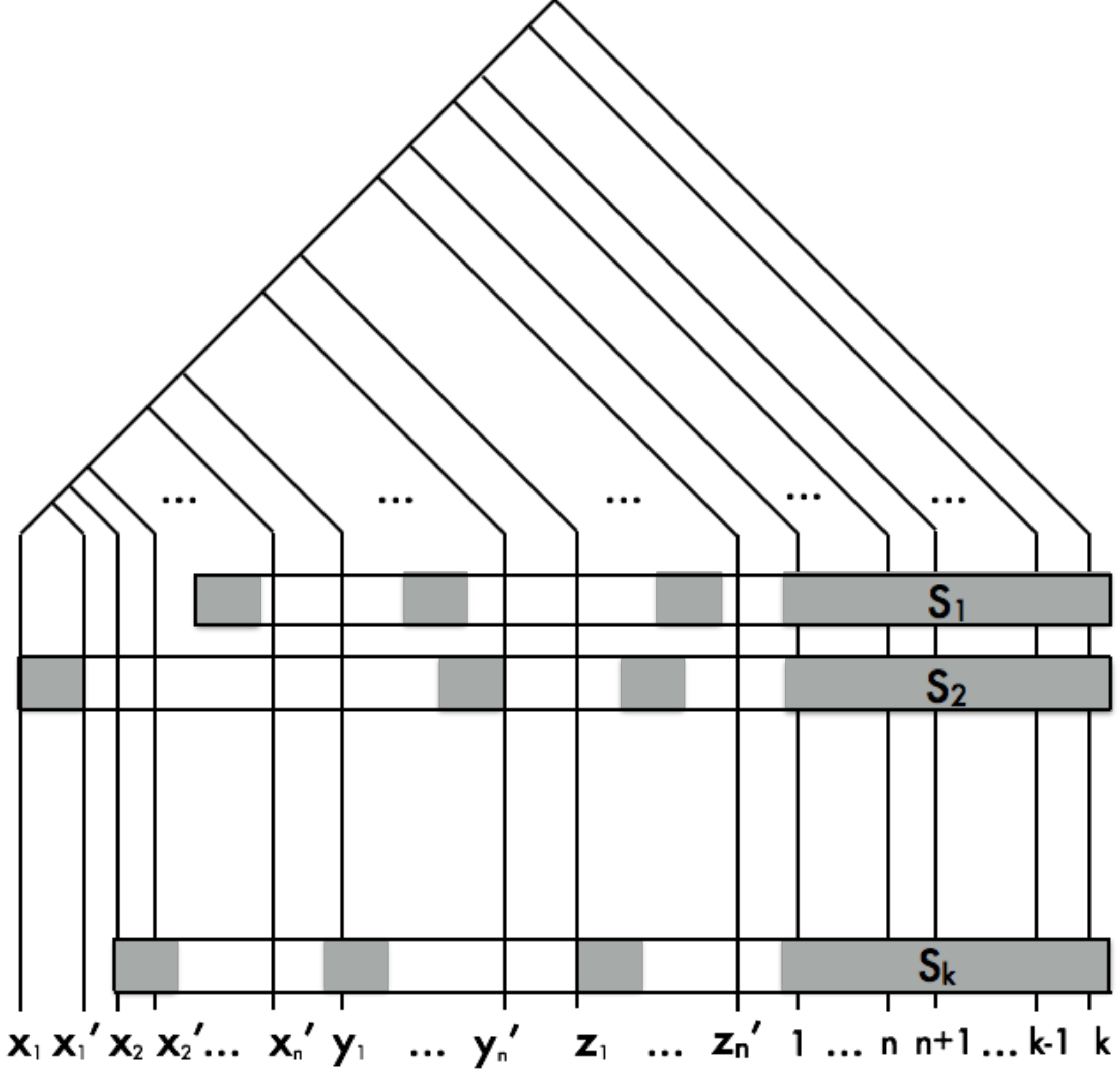}
    \hspace{0.5in}
    \includegraphics[width=2.75in]{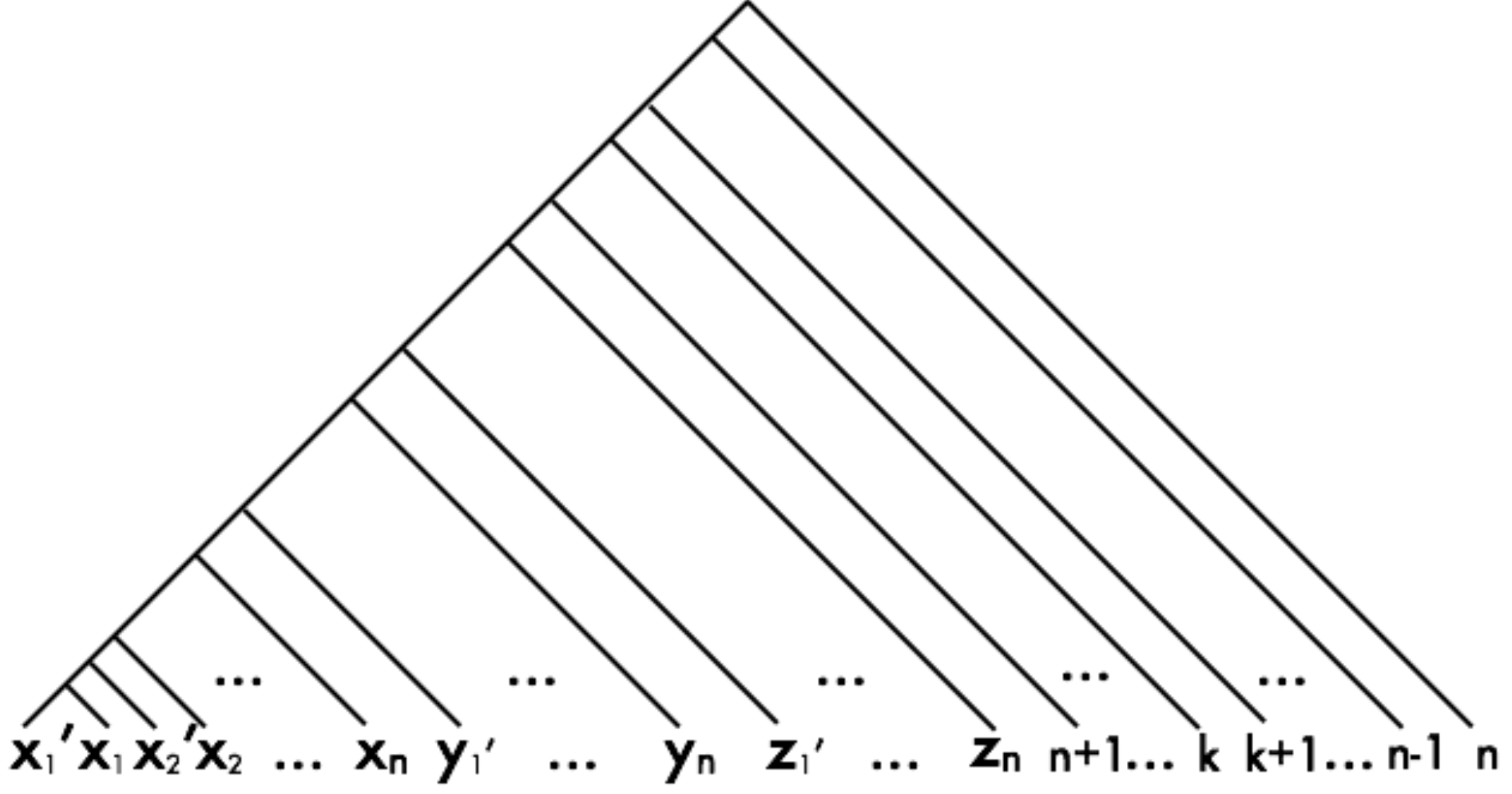}
    \caption{We encode an instance of 3 Dimensional Matching, ${\cal S} = \{S_1,S_2,\ldots,S_k\}$ by linking together gadgets for every input subset $S_i = (x_i,y_i,z_i)$.  An exact matching exists if and only if each of the element threads flip (i.e. $x_i$ and $x'_i$ switch positions, indicating that a subset containing $x_i$ was chosen) and the counter threads move exactly $n$ times, indicating exactly $n$ subsets were chosen.}
    \label{fig:reduction}
\end{figure}

That is, can we find $n$ subsets of $S$ such that all the elements in $X$, $Y$, and $Z$ are ``hit'' or  ``covered'' exactly once.
The intuition behind our construction is that we will encode each element of $X \cup Y \cup Z$ using two leaves in our network.  The leaves will ``flip'' their order when a triple their element belongs is chosen for the matching (see Figure~\ref{fig:gadget}).  To make sure that elements are only chosen once, we also add $k$ leaves which permute cyclically each time a triple is chosen. Our network consists of a gadget for each triple in the instance of 3DM.  Our tree keeps track that each element of $X \cup Y \cup Z$ belongs to a triple chosen for the matching and that no more than $n$ triples are chosen.  See Figure~\ref{fig:reduction}.

{\em Proof:}  Let sets $X$, $Y$, $Z$ with $|X|=|Y|=|Z| = n$
and $k$ subsets $S=\{S_1,\ldots,S_k\}$ be an instance of 3DM.  For each instance, we construct a phylogenetic network, $N$, and tree, $T$, on $6n+k$ leaves with leaf labels
$${\cal L} =\{x_1, x'_1, x_2, x'_2, \ldots, x_n, x'_n, 
y_1, y'_1, y_2, y'_2, \ldots, y_n, y'_n,
z_1, z'_1, z_2, z'_2, \ldots, z_n, z'_n,
1,2,\ldots, k\}.$$  
where $X = \{x_1,\ldots, x_n\}$, $Y = \{y_1,\ldots, y_n\}$, and $Z = \{z_1,\ldots, z_n\}$.

We call all trees whose internal vertices form a path terminating at the root a {\em caterpillar tree} and for brevity represent the Newick format $(((((a_1,a_2),a_3),a_4),\ldots a_n)$ as $(a_1,a_2,a_3,\ldots, a_n)$.
Define the tree $T$ as the caterpillar tree:
$$(x'_1, x_1, \ldots, x'_n, x_n, y'_1, y_1, \ldots, y'_n, y_n,
z'_1, z_1, \ldots, z'_n, z_n,
n+1,n+2, \ldots, k, 1, 2, \ldots, n-1, n)$$.

We construct the network, $N$, as follows:  The upper (or `tree') part of the network is the caterpillar tree:
$$
(x_1, x'_1, \ldots, x_n, x'_n, y_1, y'_1, \ldots, y_n, y'_n,
z_1, z'_1, \ldots, z_n, z'_n,
1,2,\ldots, k).$$
Note that these are the leaves in order, while the tree has each pair of leaves corresponding to an element (e.g. $x_i$ and $x'_i$) flipped and the remaining leaves $1,\ldots, k$ shifted $n$ positions to the left.  
The lower (or ``tangled") part of the network extends the paths to the leaves and also contains additional arcs, which together form gadgets corresponding to the subsets in $S$. Figure~\ref{fig:gadget} shows the gadget corresponding to $S_i = (x_i,y_i,z_i)$.  Each gadget is a directed acyclic graph, and the combination of the gadgets with the upper tree yields a directed acyclic graph. By construction of the gadget, there are two possible tree paths through it that give a tree-based network: 
\begin{enumerate}
    \item the tree paths do not change, or
    \item $x$ and $x'$ flip, $y$ and $y'$ flip, $z$ and $z'$ flip, and the elements $1,2,\ldots, k$ cycle one position to the left.
\end{enumerate}
The first option encodes that this subset $S_i$ is not chosen as part of the matching $M$ while the second option encodes that this subset $S_i$ is chosen as part of the matching $M$.

For each set, $S_i \in S$, we include a gadget encoding it in the lower part of the network (see Figure~\ref{fig:reduction}).  

\begin{theorem}
Let $N$ be a rooted, binary, phylogenetic network on $X$ and $T$ a rooted, binary phylogenetic tree on ${\cal L}$. Then it is NP-hard to decide if $N$ is based on $T$.
\label{thm:NP}
\end{theorem}

This reduction from an instance of 3DM to our question takes polynomial number of steps.  It only remains to show that our network $N$ is based on $T$ if and only if the 3DM instance $S$ has a matching:

First, we assume that $N$ is based on $T$ and show that 3DM instance $S$ has a matching:
Since $N$ is based on $T$, we must have that the right hand side of the gadget ends with 
$n+1, n+2, \ldots, k, 1, \ldots, n-1, n$. Since these only move when a gadget is used, we have that exactly $n$ gadgets were used in the final configuration.  Call the set of subsets $S_i$ corresponding to those gadgets, $M$.  We also have 
that every element leaf-pair of $X$, $Y$, and $Z$ is flipped.
By construction of the gadget, if $x_i$ flipped in gadget $i$, then the corresponding $y_i$ and $z_i$ of that gadget flipped.  If some $x'$ was flipped multiple times, then it must have been involved in multiple triples that were selected.  Since only $n$ triples were selected, this would mean that $<n$ elements of $X$ were flipped, contradicting the ending positions of $X$.  Thus, $M$ contains $n$ triples and each element of $X$ is contained in a triple of $M$.  By similar argument, $M$ also contains all elements of $Y$ and $Z$ yielding the desired matching.

Next, assume that the 3DM instance $S$ has a matching, $M$, and show that $N$ is based on $T$:  
To show that $N$ is based on $T$, we use the triples in the matching $M$ to show the embedding of $T$ into $N$.
For each triple in $M$, follow the "flipped" option through the corresponding gadget and choose the path that shifts the right hand paths one to the left.  For each triple not in $M$, follow the ``unflipped" option and do not shift the right hand paths. By construction, each gadget used will flip the corresponding triple $(x,y,z)$ and shift the right hand strands.  Since each element of $X$ ($Y$ and $Z$ respectively), occurs in exactly one triple and thus one gadget, all will be flipped and correspond to the final ordering of the leaves in $T$.  Since there are $n$ triples in the matching, the right hand strands will be in the correct position.  We note that by construction of the gadgets, each non-tree edge connects to tree-based edges.  Thus, if $S$ has a matching then $N$ is based on $T$.\qed

\subsection{Parametrized Algorithm}

Our final result shows that the difficulty of determining that a network $N$ is based on $T$ can be parametrized by $l$, where $l$ is $|{\cal L}|-2-y$ where $y$ is the number of $Y$-vertices in the input network $N$.  That is, 

\begin{theorem}
    Let $N$ be a rooted, binary, phylogenetic network and $T$ be a rooted, binary, phylogenetic tree both on leaf set ${\cal L}$.  Let $y$ be the number of $Y$-vertices in $N$ such that $l = |{\cal L}| - 2 -y$.  Then there exists constant $c$ and function $f$ such that $N$ being based on $T$ can be determined in time 
    $O(f(l)\cdot m^c)$
    where $f(l)$ does not depend on $m = |{\cal L}|$.
    \label{theorem:fpt}
\end{theorem}

{\em Proof:}  Let $N$, $T$ and $l$ be as in the hypothesis.  We first apply the $\lambda$ and $Y$ reduction rules of Francis and Steel \cite{francis2015}. This can be done in polynomial time in $|{\cal L}|$ and results in $\leq l$ undecided $Y$ vertices.  For each one, try both possible resolutions.  This yields $\leq 2^{l}$ possible trees.  Compare each resolved tree to $T$.  If one matches, output 'yes'.  If none matches, output 'no'.  Comparing two trees can be done in time linear in the leaves \cite{day1985}.  Since there are $\leq 2^{l}$ possible trees, the running time is bounded by $2^{l}\cdot m ^c$ for some constant $c$.
\qed

\section{Acknowledgments}

We would like to thank the American Museum of Natural History and the CUNY Advanced Science Research Center for hosting us for several meetings.  This work was funded by a Research Experience for Undergraduates (REU) grant from 
the US National Science Foundation (\#1461094 to St.~John and Owen).
{
\small
\bibliographystyle{plain}
\bibliography{phylo}
}

\end{document}